\documentclass[aps,prd,twocolumn,showpacs,superscriptaddress,groupedaddress]{revtex4-2}
\usepackage{graphicx}  
\usepackage{dcolumn}   
\usepackage{bm}        
\usepackage{amssymb}   
\usepackage{amsmath}
\usepackage{xcolor}
\usepackage{amsthm}
\usepackage{enumitem}
\usepackage{comment}
\newcommand{\be}{\begin{equation}}\newcommand{\ee}{\end{equation}}
\newcommand{\bea}{\begin{eqnarray}}\newcommand{\eea}{\end{eqnarray}}
\newcommand{\brr}{\begin{array}}\newcommand{\err}{\end{array}}
\newcommand{\bit}{\begin{itemize}}\newcommand{\eit}{\end{itemize}}
\newcommand{\ben}{\begin{enumerate}}\newcommand{\een}{\end{enumerate}}

\newcommand{\ba}{\begin{array}}
\newcommand{\ea}{\end{array}}

\definecolor{darkred}{rgb}{.8,0,0}

\definecolor{darkblue}{rgb}{0,0,.7}

\def\lf{\left}

\def\non{\nonumber}\def\pa{\partial}

\def\ri{\right}
\def\al{\alpha}\def\Ga{\Gamma}

\def\si{\sigma}

\def\1{{_{1}}}\def\2{{_{2}}}

\def\noHe0{:\;\!\!\;\!\!:H_e(0):\;\!\!\;\!\!:}
\def\noHm0{:\;\!\!\;\!\!:H_\mu(0):\;\!\!\;\!\!:}

\def\lf{\left}

\def\non{\nonumber}
\def\pa{\partial}

\def\ri{\right}

\def\al{\alpha}
\def\Ga{\Gamma}

\def\si{\sigma}

\def\1{{_{1}}}\def\2{{_{2}}}

\begin{document}
\title{Weak equivalence principle violation for mixed scalar particles}

\author{Massimo Blasone}
\email{blasone@sa.infn.it}
\affiliation{Dipartimento di Fisica, Universit\`a di Salerno, Via Giovanni Paolo II, 132 I-84084 Fisciano (SA), Italy.}
\affiliation{INFN, Sezione di Napoli, Gruppo collegato di Salerno, Italy}

\author{Petr Jizba}
\email{p.jizba@fjfi.cvut.cz}
\affiliation{FNSPE,
Czech Technical University in Prague, B\v{r}ehov\'{a} 7, 115 19, Prague, Czech Republic}
\affiliation{ITP, Freie Universit\"{a}t Berlin, Arnimallee 14, D-14195 Berlin, Germany}

\author{Gaetano Lambiase}
\email{lambiase@sa.infn.it}
\affiliation{Dipartimento di Fisica, Universit\`a di Salerno, Via Giovanni Paolo II, 132 I-84084 Fisciano (SA), Italy}
\affiliation{INFN, Sezione di Napoli, Gruppo collegato di Salerno, Italy}

\author{Luciano Petruzziello}
\email{lupetruzziello@unisa.it}
\affiliation{Dipartimento di Ingegneria, Universit\`a di Salerno, Via Giovanni Paolo II, 132 I-84084 Fisciano (SA), Italy}
\affiliation{INFN, Sezione di Napoli, Gruppo collegato di Salerno, Italy}
\affiliation{Institut f\"ur Theoretische Physik, Albert-Einstein-Allee 11, Universit\"at Ulm, 89069 Ulm, Germany}

\date{\today}

\begin{abstract}

We investigate the non-relativistic limit of the Klein--Gordon equation for mixed scalar particles and 
show that, in this regime, one unavoidably arrives at redefining the
particle's inertial mass. This happens because, in contrast to the case when mixing is absent, the antiparticle sector contribution cannot be neglected for particles with definite flavor.
To clearly demonstrate this feature, we adopt the Feshbach--Villars formalism for Klein--Gordon particles.
Furthermore, within the same framework, we also demonstrate that, in the presence of a weak gravitational field, the mass parameter that couples to gravity (gravitational mass) does not match the effective inertial mass. This, in turn, implies a violation of the weak equivalence principle. Finally, we prove that the Bargmann's superselection rule, which prohibits oscillating particles on the basis of the Galilean transformation, is incompatible with the non-relativistic limit of the Lorentz transformation and hence does not collide with the results obtained.

\end{abstract}

\vskip -1.0 truecm

\maketitle
\section{Introduction}

Together with the notion of general covariance, the equivalence principle represents a cornerstone of general relativity. Since its inception more than one century ago, there have been a number of distinct interpretations of the equivalence principle, each with its own set of assumptions and scope of use~\cite{ep,ep2,ep3,wep6,ep4,ep5}. The simplest version of the equivalence principle, the so-called weak equivalence principle (WEP), asserts that non-inertial effects caused by acceleration are indistinguishable from the effects of an external gravitational field. This is encoded in the identity between the inertial and gravitational mass of test bodies subject to gravity. 

Among the many theoretical and experimental 
efforts that have been done in connection with 
the WEP (a comprehensive review of recent developments can be found in Ref.~\cite{wep6}), particularly prominent are those based on particle mixing, because these can also shed light on the validity of WEP in the quantum regime. Specifically, the investigations of WEP in neutrino physics have been a driving force behind a large number of studies, cf. Refs.~\cite{sck,nmf,sah,ash,bha,nprd,ijmpd,gn,hl} and references therein. 
Along these lines, the primary focus has been on ultra-relativistic neutrinos, as they are the most pertinent from a phenomenological standpoint. However, recently it has been pointed out~\cite{BJLP} that, in a non-relativistic setting, it is simple to differentiate between inertial mass $m_i$ and gravitational neutrino mass $m_g$ in the weak interaction basis (i.e., flavor basis).
In other words, for low energy neutrinos the violation of  WEP becomes manifest. 

In the case of neutrinos, the mismatch between $m_i$  and $m_g$ can be attributed to the unavoidable presence of flavor mixing. In particular, when performing the non-relativistic limit, one has to simultaneously deal with large and small bispinor components  
(e.g., $\psi^L$ and $\psi^S$), which in the case of mixing are both comparably important. In fact, to identify the low-energy inertial mass $m_i$, one has to work interchangeably with both small and large components because these are interlocked at all energy scales.  On the other hand, when the conventional minimal coupling to gravity is considered in the weak-field approximation, the gravitational potential couples directly to the original flavor masses,
allowing them to be interpreted as equivalent to gravitational masses.
Since the gravitational mass does not undergo the same redefinition as the inertial one, the violation of WEP arises.

At this stage, one may wonder if a similar violation of WEP would be also observed for spin-0 particles, since particle mixing is not exclusive to neutrino physics, but it is also found for weakly interacting mesons.
The most important mixing phenomena in this context are represented by the oscillations of neutral kaons $K^0\rightleftharpoons \bar{K}^0$, neutral and strange $B$ mesons, $B^0\rightleftharpoons \bar{B}^0$ and $B_s^0\rightleftharpoons \bar{B}_s^0$, respectively, and neutral $D$ mesons $D^0\rightleftharpoons \bar{D}^0$. Although the last example has been definitively established only recently~\cite{dmes}, the other meson oscillations have been known for a long time. In particular, the strange $B$ meson has attracted a significant attention over the years, as its phenomenology is considered to be relevant for understanding the asymmetry of matter and antimatter in the visible universe~\cite{mam,mam2}.

In this paper, we aim to expand the results of Ref.~\cite{BJLP} by studying the non-relativistic limit of mixed scalar particles and determine if the difference between $m_i$ and $m_g$ observed in~\cite{BJLP} for mixed Dirac fermions holds also for spinless particles.
To keep our considerations as close as possible to the spin-$\frac{1}{2}$ case, we will resort to the Feshbach--Villars (FV) representation of a Klein-Gordon particle~\cite{FV,JK}. The FV representation allows to reformulate  the Klein--Gordon equation in a Schr\"odinger-like form (thereby employing first-order time derivatives, as for the Dirac equation), where 
the ensuing wave function has a two-component form, which reflects the presence of particles and antiparticles~\cite{FV}. It should also be stressed that, in a non-relativistic limit,
the positive-energy plane-wave solutions of the FV equation have the upper component much larger than the lower components~\cite{JK}. Similarly, for the negative-energy plane wave solutions one gets that the lower component is much larger. 
The analogous situation holds also for Dirac wave functions, with the only difference that, in the latter case, one must also consider helicity components. 

In contrast to Dirac fermions where gravity couples through minimal coupling (with the Fock--Kondratenko connection), 
there is not yet a standard theory of massive spinless bosons in curved spacetime~\cite{BD}.  
In our considerations, we will employ the commonly used  conformal coupling, which, among others, has a correct quasiclassical limit, avoids a tachyonic behavior and allows for a straightforward translation in terms of the FV representation~\cite{BD,fvgrav,GP}.

For the sake of consistency, we supplement our discussion by examining Bargmann's superselection rules~\cite{Barg}. Specifically, we demonstrate why superpositions of states with different masses are not problematic in the non-relativistic limit of relativistic quantum mechanics, unlike the case in which one instead starts from Galilean (rather than Lorentz) invariance.

The paper is organized as follows: in Sec.~II, we briefly review the non-relativistic limit for mixed neutrinos in the weak interaction basis. In Section~III, we study in detail the non-relativistic limit of the FV equation for mixed scalar particles and show that, in this framework, one inevitably comes across a non-trivial correction to the initial inertial mass, which leads to a low-energy effective inertial mass $m_i$. 
In addition, if a weak gravitational field is present, we show
%
%
that the corresponding gravitational mass $m_g$ does not undergo the same redefinition as the inertial mass, hence $m_i \neq m_g$, which is a direct signature of WEP violation.  
In Section~IV, we briefly comment on the violation of the Bargmann's superselection rules in connection with the non-relativistic limit of a relativistic theory with mixed particles.  
Concluding remarks and generalizations are proposed in Section~V.


\section{Non-relativistic mixed neutrinos}

In this Section, we briefly summarize the results of Ref.~\cite{wep6} regarding the violation of WEP for mixed neutrinos.
To this end, we start from the two-flavor Dirac equation associated with neutrinos $\nu_e$ and $\nu_\mu$. In a compact notation, this reads
\be
\lf(i\gamma'^\al\partial_\al \ - \ \mathbb{M}\ri)\Psi \ = \ 0\,,
\label{deq}
\ee
where $\gamma'^\al$ is the $8\times 8$ matrix $\mathbb{I}_{2\times 2}\otimes\gamma^\al$,
$\mathbb{M}$ is the $8\times 8$ non-diagonal mass matrix, which in the $4\times 4$ block formalism is given by
\be\label{M}
\mathbb{M} \ = \ \begin{pmatrix}m_e & m_{e\mu} \\m_{e\mu} & m_\mu\end{pmatrix}\!,
\ee
(note that $m_{e\mu} = m_{\mu e}$), whilst the wave-function $\Psi$ is a short-hand notation for the neutrino doublet
\be\label{sp}
\Psi\ = \ \begin{pmatrix}\psi_e\\\psi_\mu\end{pmatrix}\!.
\ee
At this stage, one can choose to work with the electron neutrino only, as the implications for the muon neutrino can be derived by simply swapping the subscripts
$e\leftrightarrow\mu$. So, for positive energy solutions, we obtain  from Eq.~\eqref{deq} two coupled algebraic equations in momentum space:
\bea\non
\lf(E_e-m_e\ri)\varphi_e \ - \ {\boldsymbol{\si}}\cdot{\boldsymbol{p}}\,\chi_e &=& m_{e\mu}\varphi_\mu\,,\\[2mm]
{\boldsymbol{\si}}\cdot{\boldsymbol{p}}\,\varphi_e \ - \ \lf(E_e+m_e\ri)\chi_e &=& m_{e\mu}\chi_\mu\, ,
\label{en1}
\eea
with $\varphi_{e,\mu}$ and $\chi_{e,\mu}$ denoting the ``large'' (upper)  and ``small'' (lower) bispinor components. To consider the non-relativistic limit, one should bear in mind that the dominant contribution to the energy comes from the rest mass. Hence, in Eqs.~(\ref{en1}) we can define the non-relativistic energy $E^{N\!R}_{e} \equiv E_{e} - m_{e}$,
and write~(\ref{en1}) as
\bea\label{e2a} \non
\mbox{\hspace{-4mm}}E^{N\!R}_{e}{\varphi}_e \ - \ {\boldsymbol{\si}}\cdot{\boldsymbol{p}}\ \! \,{\chi}_e &=& m_{e\mu}\ \!{\varphi}_\mu\,,\\[2mm]\label{e2}
{\boldsymbol{\si}}\cdot{\boldsymbol{p}}\ \!\,{\varphi}_e\ - \ 2m_e{\chi}_e &=& m_{e\mu}\ \!{\chi}_\mu\,.
\label{5.cc}
\eea
Analogous equations hold for $e\leftrightarrow\mu$.  

If there was no mixing, the small component $\chi$ would be negligible compared to the large component $\varphi$.
By taking particle mixing into account, 
the small component $\chi_e$ will still remain much smaller than $\varphi_e$, provided a small admixture of the large component $\varphi_{\mu}$ is included (a similar statement holds also for  $e\leftrightarrow\mu$). This can be seen as follows: we first eliminate $\chi_\mu$  in the second equation in (\ref{5.cc}) 
by inserting $\chi_\mu$ from the analogous equation where $e$ and $\mu$ are exchanged. With this, the second equation in (\ref{5.cc})  can be cast in the form
\begin{eqnarray}
\chi_e \ = \ \frac{{\boldsymbol{\si}}\cdot{\boldsymbol{p}}}{2 m_e} \ \! \varphi_e\ - \ \frac{m_{e\mu}}{4m_e m_{\mu}}\ \! {\boldsymbol{\si}}\cdot{\boldsymbol{p}} \ \! \varphi_{\mu}\ + \ \omega \ \!\chi_e\, ,
\end{eqnarray}
or equivalently
\begin{eqnarray}
\chi_e \ = \ \frac{{\boldsymbol{\si}}\cdot{\boldsymbol{p}}}{(1-\omega)2 m_e} \ \! \varphi_e\ - \ \frac{m_{e\mu} \ \! {\boldsymbol{\si}}\cdot{\boldsymbol{p}}}{(1- \omega)4m_e m_{\mu}} \ \! \varphi_{\mu}\, ,
\label{7.cv}
\end{eqnarray}
where
\be\label{exppar}
\omega \ = \ \frac{m_{e\mu}^2}{4m_em_\mu}\,.
\ee
Thus, by assuming that $m_e \le m_\mu$, we get to the order $\mathcal{O}(\omega)$ that
\begin{eqnarray}
|\!| \chi_e |\!|_2 &\le & \frac{|\boldsymbol{p}|}{2}\left( \frac{|\!|\varphi_{e} |\!|_2}{m_e} + \frac{\sqrt{\omega}}{\sqrt{m_e m_{\mu}}}\ \! |\!|\varphi_{\mu} |\!|_2\right)\nonumber \\[2mm]
&\le & \frac{|\boldsymbol{p}|}{2 m_e}\left( |\!|\varphi_{e} |\!|_2 + \sqrt{\omega} |\!|\varphi_{\mu} |\!|_2\right)\, ,
\label{9.cc}
\end{eqnarray}
and similarly
\begin{eqnarray}
|\!| \chi_\mu |\!|_2 &\le &
\frac{|\boldsymbol{p}|}{2 \sqrt{m_e m_{\mu}}}\left( |\!|\varphi_{\mu} |\!|_2 + \sqrt{\omega} |\!|\varphi_{e} |\!|_2\right)\, ,
\label{10cc}
\end{eqnarray}
where $|\!| \ldots |\!|_2$ denotes the $\ell_2$-norm. In particular, when $\sqrt{\omega} |\!|\varphi_{\mu} |\!|_2 \lesssim |\!|\varphi_{e} |\!|_2$ and $\sqrt{\omega} |\!|\varphi_{e} |\!|_2 \lesssim |\!|\varphi_{\mu} |\!|_2$, we have that in the non-relativistic limit (i.e., when $|\boldsymbol{p}| \ll m_e$) the large components are much larger than their respective small components.
%
Note that the relativistic regime enters at scales where $|\boldsymbol{p}| \gg \sqrt{m_e m_{\mu}}$, which is also in agreement with QFT considerations~\cite{BJV}.


By substituting $\chi_e$ from (\ref{7.cv}) into the first equation in (\ref{e2a}), we eliminate the small component from the equation and obtain  
%
%
%
%
%
\bea\non
E^{N\!R}_{e}\ \!\varphi_e  &=&  \frac{\boldsymbol{p}^2}{2m_e(1-\omega)}\ \!\varphi_e \\[2mm]
&+&  \lf[m_{e\mu}\ - \ \frac{m_{e\mu}}{2m_e}\ \! \frac{1}{(1-\omega)} \ \! \frac{\boldsymbol{p}^2}{2m_\mu}\ri]\ \!\varphi_\mu \nonumber \\[2mm]
&=&  \frac{\boldsymbol{p}^2}{2m_{i,e}}\ \!\varphi_e \ + \ m_{e\mu}\left[1 \ - \ \frac{\boldsymbol{p}^2}{4m_e m_{i,\mu}}\right]\varphi_\mu \, 
.~~~~~~~\label{fe3}
\eea
%
%
where we have defined the effective inertial mass $m_{i,e}  =   m_{e}\left(1-{\omega}\right)$ (and similarly for $m_{i,{\mu}}$).
 In Ref.~\cite{BJLP}, the same result was obtained 
by means of the iteration method. An analogous relation holds also for $e\leftrightarrow\mu$. 

Equation~(\ref{fe3}) is the sought non-relativistic limit of the Dirac equation for a mixed electron neutrino. 
%
If $m_{e\mu}$ is equal to zero, the  equations for $\varphi_e$ and $\varphi_\mu$ become independent, resulting in free electron and muon neutrinos with masses $m_e$ and $m_\mu$, respectively.
The presence of the amplitude in square brackets in the r.h.s. of \eqref{fe3}, however, creates a connection between the two flavor neutrinos, implying that one flavor might ``leak'' into the other.
In fact, such an amplitude is nothing but the ``flip-flop'' amplitude of a two-state system~\cite{Feyn}. 
The amplitude modulus is manifestly invariant under the exchange of flavors $e \leftrightarrow \mu$, thus reflecting the principle of detailed balance in the oscillation phenomenon.
From~(\ref{fe3}), it is evident that the inertial mass $m_i$ appearing in the kinetic term must be re-scaled by a factor $1-\omega$, which  reduces to unity only when mixing is removed, \emph{i.e.}, when $m_{e\mu}\to0$.

The root cause of the violation of the weak equivalence principle for mixed particles can be traced back to the previous observation. Indeed, 
the weak equivalence principle states that inertial and gravitational mass are equal, and it is not difficult to check that,
when a gravitational potential is switched on, then the gravitational mass is not redefined in the same way as the inertial mass. To see this, it is convenient to study the Dirac equation in the weak-field regime of the Schwarzschild solution in isotropic coordinates. The line element in this case is expressed as follows:
\be\label{g}
ds^2   =    \lf(1+2\,\phi\,\ri)dt^2 \ -  \lf(1-2\,\phi\ri)\lf(dx^2+dy^2+dz^2\ri)\,,
\ee
with $\phi=-GM/|\boldsymbol{x}|$ being the Newtonian potential. Now, the Dirac equation must be rephrased to take into account the presence of gravity. This is done by means of the Fock--Kondratenko connection $\Gamma_\mu$, which replaces the standard derivatives appearing in~\eqref{deq} with covariant ones $\pa_\mu\to D_\mu=\pa_\mu+\Ga_\mu$. 
%
%
%
In the simplest case of a slowly varying potential (\emph{i.e.}, $\pa_i\phi\approx0$ with $i=x,y,z$), Eq.~\eqref{fe3} is modified into (cf.  Ref.~\cite{BJLP})
\bea\non
E^{N\!R}_{e}\ \!\varphi_e \ &=& \ \lf[\frac{\boldsymbol{p}^2}{2m_e(1-\omega)} \ + \ m_e\phi\ri] \varphi_e \\[2mm]
&+& \ V\lf(\boldsymbol{p}^2,\phi\ri)\ \!\varphi_\mu \ + \ \mathcal{O}(\boldsymbol{p}\phi)\,,\label{neutfinal}
\eea
where $\mathcal{O}(\boldsymbol{p}\phi)$ represent post-Newtonian corrections and the details of $V(\boldsymbol{p}^2,\phi)$ are not relevant for the purpose of the current analysis. In fact, the most important term in the above equation is the term representing the Newtonian coupling to the external potential, from which the expression for the gravitational mass $m_g=m_e$ can be unambiguously identified. Therefore, since $m_g$ is not redefined in the same way as the inertial mass, we can conclude that $m_i \neq m_g$ for mixed particles, which implies the violation of the weak form of the equivalence principle.
A similar equation also holds when $e \leftrightarrow \mu$.

Let us stress in passing that the results of the WEP violation were obtained by considering  mixed neutrinos in the flavor basis and with only two generations. This approach is not applicable in the mass basis, because the non-relativistic limit and the mixing transformations are not interchangeable in this case~\cite{BJLP}.

In the following Section, we will observe that the violation of WEP is not exclusive to mixed Dirac fields, but also occurs for mixed scalar fields.


\section{Mixed Klein--Gordon particles \label{Section II}}

\subsection{General setup \label{SubSect IIa}}

To keep our discussion as close as possible to the Dirac equation treatment from the previous Section,
we start from the observation that the Klein--Gordon equation for a free spinless
particle can be rewritten in a Schr\"{o}dinger-like form --- the so-called
Feshbach--Villars representation (FVR), cf. e.g., Refs.~\cite{FV,JK}, as
\begin{eqnarray}
i \partial_t \Phi \ = \ H_{FV}(\hat{{\boldsymbol{p}}}) \Phi\, ,
\label{FV0}
\end{eqnarray}
where $\hat{{\boldsymbol{p}}} = -i \nabla_{\!{\boldsymbol{x}}}$. The wave function $\Phi$ is
a two-component object
\begin{eqnarray}
\Phi \ = \ \left(
             \begin{array}{c}
               \zeta \\[1mm]
               \bar{\zeta }\\
             \end{array}
           \right),
\end{eqnarray}
and the Hamiltonian operator is a $2\times 2$ Hermitian matrix
\begin{eqnarray}
H_{FV}(\hat{{\boldsymbol{p}}}) \ = \ \left(\sigma_3 + i \sigma_2 \right) \frac{\hat{{\boldsymbol{p}}}^2}{2m} + \sigma_3 m c^2\, .
\label{FV1}
\end{eqnarray}
Note that the components $\zeta$ and $\bar{\zeta}$ can be represented explicitly as
\begin{eqnarray}
&&\zeta \ = \ \frac{1}{\sqrt{2}} \left(\psi \ - \ \frac{1}{im c^2} \frac{\partial \psi}{\partial t}  \right), \nonumber \\[2mm]
&&\bar{\zeta} \ = \ \frac{1}{\sqrt{2}} \left(\psi \ + \ \frac{1}{im c^2} \frac{\partial \psi}{\partial t}  \right),
\end{eqnarray}
where $\psi$ is a Klein--Gordon wave function satisfying
\begin{eqnarray}
\left(\square \ + \  m^2c^2   \right) \psi \ = \ 0\, .
\end{eqnarray}
We might note in passing that the charge-conjugated wave function has the form~\cite{FV}
\begin{eqnarray}
\Phi_c \ = \ \sigma_{1} \Phi_c^*
 \ = \  \left(
             \begin{array}{c}
               \bar{\zeta }^* \\[1mm]
                \zeta^*\\
             \end{array}
           \right). 
 \end{eqnarray}
The two-component form of the FV wave function thus indicates the existence of both particles and their antiparticles.

Within the FVR, the two-flavor mixing of scalar particles can be formulated in a similar manner to that of neutrino mixing. Namely, we first  write two
decoupled Eqs.~(\ref{FV0}) for different masses $m_1$ and $m_2$ (as in the case of
the mass basis). We then rotate the ensuing diagonal mass matrix to a flavor basis where
the mass matrix acquires also off-diagonal terms, so that
\begin{eqnarray}
\left(
  \begin{array}{cc}
    m_1 & 0 \\
    0 & m_2 \\
  \end{array}
\right) \ \mapsto \ \left(
                      \begin{array}{cc}
                        m_{_{\rm{I}}} & m_{_{\rm{I,II}}} \\
                        m_{_{\rm{I,II}}} & m_{_{\rm{II}}} \\
                      \end{array}
                    \right) \ \equiv \ \mathbb{M}\, ,
                    \label{20.cv}
\end{eqnarray}
and $(\Phi_{1}, \Phi_{2})^{{{T}}} \mapsto (\Phi_{_{\rm{I}}}, \Phi_{_{\rm{II}}})^{{{T}}}$.

By formally changing $m$ to $\mathbb{M}$ in the Hamiltonian~(\ref{FV1}), we can explicitly write the two FVR equations for mixed particles in the form ($c= 1$)
\begin{eqnarray}
i \partial_t \Phi_{_{\rm{I}}} &=&  \left(\sigma_3 + i \sigma_2 \right) \frac{\hat{{\boldsymbol{p}}}^2}{2D}
\ \! \left(m_{_{\rm{II}}}\Phi_{_{\rm{I}}} - m_{_{\rm{I,II}}}\Phi_{_{\rm{II}}} \right)\nonumber \\[2mm] &&+ \ \sigma_3  \ \! \left(m_{_{\rm{I}}}\Phi_{_{\rm{I}}} +  m_{_{\rm{I,II}}}\Phi_{_{\rm{II}}}   \right)\, ,\label{B34a} \\[2mm]
i \partial_t \Phi_{_{\rm{II}}} &=&  \left(\sigma_3 + i \sigma_2 \right) \frac{\hat{{\boldsymbol{p}}}^2}{2D}
\ \! \left(m_{_{\rm{I}}}\Phi_{_{\rm{II}}} - m_{_{\rm{I,II}}}\Phi_{_{\rm{I}}} \right)\nonumber \\[2mm] &&+ \ \sigma_3  \ \! \left(m_{_{\rm{II}}}\Phi_{_{\rm{II}}} +  m_{_{\rm{I,II}}}\Phi_{_{\rm{I}}}   \right)\, ,
\label{B34b}
\end{eqnarray}
where $D$ represents the (flavor basis) mass matrix determinant
\begin{eqnarray}
D \ = \ m_{_{\rm{I}}}m_{_{\rm{II}}} - m_{_{\rm{I,II}}}^2 \ \equiv \ m_{_{\rm{I}}}m_{_{\rm{II}}} (1- \bar{\omega})\, .
\end{eqnarray}
Here, $\bar{\omega} = m_{_{\rm{I,II}}}^2/(m_{_{\rm{I}}}m_{_{\rm{II}}})$ is an analogue of $\omega$ from Eq.~(\ref{exppar}).

\subsection{Non-relativistic limit \label{SubSect IIb}}

Let us now focus on Eq.~(\ref{B34a}), since the following reasoning for $\Phi_{_{\rm{I}}}$
can be easily repeated also for $\Phi_{_{\rm{II}}}$  via the exchange of subscripts ${\rm{I}} \leftrightarrow {\rm{II}}$. In
momentum representation, the positive-energy wave functions satisfy the algebraic equations
\begin{eqnarray}
E_{_{\rm{I}}} \Phi_{_{\rm{I}}}^L &=& \frac{{{\boldsymbol{p}}}^2}{2D}\left(m_{_{\rm{II}}}\Phi_{_{\rm{I}}}^L  + m_{_{\rm{II}}}\Phi_{_{\rm{I}}}^S - m_{_{\rm{I,II}}}\Phi_{_{\rm{II}}}^L - m_{_{\rm{I,II}}}\Phi_{_{\rm{II}}}^S \right)\nonumber \\[2mm]
&&+ \ m_{_{\rm{I}}} \Phi_{_{\rm{I}}}^L + m_{_{\rm{I,II}}} \Phi_{_{\rm{II}}}^L\, , \label{B36.a}\\[2mm]
E_{_{\rm{I}}} \Phi_{_{\rm{I}}}^S &=& \frac{{{\boldsymbol{p}}}^2}{2D}\left(m_{_{\rm{I,II}}}\Phi_{_{\rm{II}}}^L  + m_{_{\rm{I,II}}}\Phi_{_{\rm{II}}}^S - m_{_{\rm{II}}}\Phi_{_{\rm{I}}}^L - m_{_{\rm{II}}}\Phi_{_{\rm{I}}}^S \right)\nonumber \\[2mm]
&&- \ m_{_{\rm{I}}} \Phi_{_{\rm{I}}}^S - m_{_{\rm{I,II}}} \Phi_{_{\rm{II}}}^S\, .
\label{B36.b}
\end{eqnarray}
The superscripts $L$ and $S$ denote the ``large'' (upper)  and ``small'' (lower) components of the FV wave functions, respectively~\cite{FV,JK}. The non-relativistic
limit of (\ref{B36.a})-(\ref{B36.b}) is now obtained along the same line as in the Dirac case, namely
\begin{eqnarray}
E^{N\!R}_{_{\rm{I}}} \Phi_{_{\rm{I}}}^L\! &=&\! \frac{{{\boldsymbol{p}}}^2}{2D}\left[m_{_{\rm{II}}}(\Phi_{_{\rm{I}}}^L  + \Phi_{_{\rm{I}}}^S) - m_{_{\rm{I,II}}}(\Phi_{_{\rm{II}}}^L + \Phi_{_{\rm{II}}}^S) \right]\nonumber \\[2mm]
&&\!+ \ m_{_{\rm{I,II}}} \Phi_{_{\rm{II}}}^L\, ,
\label{B.40.b}
\end{eqnarray}
and
\begin{eqnarray}
\Phi_{_{\rm{I}}}^S \!&=& \! \frac{{{\boldsymbol{p}}}^2}{4Dm_{_{\rm{I}}} }\left[m_{_{\rm{I,II}}}(\Phi_{_{\rm{II}}}^L  + \Phi_{_{\rm{II}}}^S) - m_{_{\rm{II}}}(\Phi_{_{\rm{I}}}^L + \Phi_{_{\rm{I}}}^S) \right]\nonumber \\[2mm]
&&\!- \ \frac{m_{_{\rm{I,II}}}}{2m_{_{\rm{I}}}} \ \!\Phi_{_{\rm{II}}}^S\, ,
\label{B.41.b}
\end{eqnarray}
where $E^{N\!R}_{_{\rm{I}}} \equiv E_{_{\rm{I}}} - m_{_{\rm{I}}}$ and $E_{_{\rm{I}}} + m_{_{\rm{I}}} \approx 2m_{_{\rm{I}}}$. 
As in the neutrino case, one can now use the non-relativistic relation
for $\Phi_{_{\rm{II}}}^S$, which reads, cf.~(\ref{B.41.b})
\begin{eqnarray}
\Phi_{_{\rm{II}}}^S \!&=& \! \frac{{{\boldsymbol{p}}}^2}{4Dm_{_{\rm{II}}} }\left[m_{_{\rm{I,II}}}(\Phi_{_{\rm{I}}}^L  + \Phi_{_{\rm{I}}}^S) - m_{_{\rm{I}}}(\Phi_{_{\rm{II}}}^L + \Phi_{_{\rm{II}}}^S) \right]\nonumber \\[2mm]
&&\!- \ \frac{m_{_{\rm{I,II}}}}{2m_{_{\rm{II}}}} \ \!\Phi_{_{\rm{I}}}^S\, ,
\label{B.41.c}
\end{eqnarray}
and from (\ref{B.41.b})-(\ref{B.41.c}) resolve $\Phi_{_{\rm{I}}}^S$ and $\Phi_{_{\rm{II}}}^S$ in terms of $\Phi_{_{\rm{I}}}^L$ and $\Phi_{_{\rm{II}}}^L$. 
On the one hand, by assuming
that $m_{_{\rm{I}}}\le m_{_{\rm{II}}}$ and $m_{_{\rm{I}}} \gg |{\boldsymbol{p}}|$, we can write for $|\Phi_{_{\rm{I}}}^S|$ up to the order $\mathcal{O}(\bar{\omega})$ 
\begin{eqnarray}
|\Phi_{_{\rm{I}}}^S| & \le & \frac{{{\boldsymbol{p}}}^2}{4} \left(\frac{|\Phi_{_{\rm{I}}}^L|}{m_{_{\rm{I}}}^2} \ + \   \frac{\sqrt{\bar{\omega}} \ \!(m_{_{\rm{I}}}+2m_{_{\rm{II}}})}{2 (m_{_{\rm{I}}}m_{_{\rm{II}}})^{3/2}} \ \! |\Phi_{_{\rm{II}}}^L|\right) \nonumber \\[2mm]
&\le & \frac{{{\boldsymbol{p}}}^2}{4m_{_{\rm{I}}}^2}\left(|\Phi_{_{\rm{I}}}^L| \ + \ \frac{3}{2}\sqrt{\bar{\omega}} \ \! |\Phi_{_{\rm{II}}}^L| \right)\, .
\label{28.cc}
\end{eqnarray}
Similarly, for $|\Phi_{_{\rm{II}}}^S|$ we get
\begin{eqnarray}
|\Phi_{_{\rm{II}}}^S| \ \le \ \frac{{{\boldsymbol{p}}}^2}{4m_{_{\rm{I}}}^2}\left(|\Phi_{_{\rm{II}}}^L| \ + \ \frac{3}{2}\sqrt{\bar{\omega}} \ \! |\Phi_{_{\rm{I}}}^L| \right)\, .
\label{29.cc}
\end{eqnarray}
The results of (\ref{28.cc}) and (\ref{29.cc}) demonstrate that the small components $\Phi_{_{\rm{I}}}^S$ and $\Phi_{_{\rm{II}}}^S$ still remain much smaller than $\Phi_{_{\rm{I}}}^L$ and $\Phi_{_{\rm{II}}}^L$, respectively, even after having introduced particle mixing.
%
%

On the other hand, by inserting the solutions for $\Phi_{_{\rm{I}}}^S$ and $\Phi_{_{\rm{II}}}^S$ back into (\ref{B.40.b}), we obtain after some algebra that
\begin{eqnarray}
E^{N\!R}_{_{\rm{I}}} \Phi_{_{\rm{I}}}^L\! &=&\!  \bar{A}(\mathbb{M})\ \! \frac{{\boldsymbol{p}}^2}{2 m_{_{\rm{I}}}}\ \! \Phi_{_{\rm{I}}}^L \ + \ \bar{B}(\mathbb{M})\ \!\Phi_{_{\rm{II}}}^L\,,
\label{28.cv}
\end{eqnarray}
where
\begin{widetext}
\begin{eqnarray}\label{Aaa}
&&\bar{A}(\mathbb{M}) \ = \   \frac{4 m_{_{\rm{I}}}^2 (4 m_{_{\rm{II}}}^2 \ + \  {\boldsymbol{p}}^2)
\ - \ 4 m_{_{\rm{I}}}m_{_{\rm{II}}}m_{_{\rm{I,II}}}^2 }{4D(4m_{_{\rm{I}}}m_{_{\rm{II}}}  -  m_{_{\rm{I,II}}}^2)  \ + \ 4(m_{_{\rm{I}}}^2  +  m_{_{\rm{II}}}^2  +  m_{_{\rm{I,II}}}^2){\boldsymbol{p}}^2  \ + \
{\boldsymbol{p}}^4}\, , \\[2mm]
\label{Bbb}
&&\bar{B}(\mathbb{M})\ = \ m_{_{\rm{I,II}}} \ - \ \frac{m_{_{\rm{I,II}}}\left(8m_{_{\rm{I}}}m_{_{\rm{II}}} - 2 m_{_{\rm{I,II}}}^2 - {\boldsymbol{p}}^2   \right)}{2m_{_{\rm{I}}} \left( 4 m_{_{\rm{II}}}^2 +  {\boldsymbol{p}}^2\right) - 2 m_{_{\rm{II}}} m_{_{\rm{I,II}}}^2 }\ \! \bar{A}(\mathbb{M}) \ \! \frac{{\boldsymbol{p}}^2}{2m_{_{\rm{I}}}}\,.
\end{eqnarray}
By employing the non-relativistic assumption $m_{_{\rm{I}}}, m_{_{\rm{II}}} \gg |{\boldsymbol{p}}|$, Eq.~(\ref{28.cv}) reduces to
\begin{eqnarray}
E^{N\!R}_{_{\rm{I}}} \Phi_{_{\rm{I}}}^L \ &=& \    \frac{{\boldsymbol{p}}^2}{2m_{_{\rm{I}}} (1-\bar{\omega})} \ \! \Phi_{_{\rm{I}}}^L
 \ + \ \left\{m_{_{\rm{I,II}}}\left[1 -   \frac{{\boldsymbol{p}}^2}{2m_{_{\rm{I}}} m_{_{\rm{II}}} (1-\bar{\omega})}\right]\right\}\Phi_{_{\rm{II}}}^L \nonumber \\[2mm] &=& \    \frac{{\boldsymbol{p}}^2}{2m_{i,{_{\rm{I}}}}} \ \! \Phi_{_{\rm{I}}}^L
  \ + \  \left\{m_{_{\rm{I,II}}}\left[1  -   \frac{{\boldsymbol{p}}^2}{2m_{_{\rm{I}}}m_{i,{_{\rm{II}}}} } \right]\right\}\Phi_{{_{\rm{II}}}}^L
\,,
\end{eqnarray}
\end{widetext}
where we have defined the effective inertial mass $m_{i,{_{\rm{I}}}}  =   m_{_{\rm{I}}}\left(1-\bar{\omega}\right)$ (and similarly for $m_{i,{_{\rm{II}}}}$). An analogous equation holds also for $I\leftrightarrow II$.
This outcome should be compared with the expression (\ref{fe3}) for  flavor neutrinos.
The extra factor 2  in (\ref{fe3}) (respective 4 in $\omega$)
is a consequence of the way how the factor appears in the kinetic versus mixing term in non-relativistic
equations (\ref{e2}) and (\ref{B.40.b})-(\ref{B.41.b}), thereby denoting a different spin content.
Such a spin-dependent behavior of the effective mass  can  also be observed for higher-spin particle states described via Bargmann--Wigner equations.

\subsection{Non-relativistic limit in presence of  gravitational field\label{SubSect IIIb}}

Let us now focus on what happens when we switch a gravitational potential on. It is not a priori evident whether the effective inertial masses $m_{i,{_{\rm{I}}}}$ and $m_{i,{_{\rm{II}}}}$ will also couple to the gravitational potential. 
To explore this issue, we will employ the conformal coupling to gravity and restrict our analysis to the weak-field metric~\eqref{g}.

When mixing is absent, one can show~\cite{fvgrav} that the form of the Klein--Gordon equation in the Feshbach--Villars representation~\eqref{FV0} gets modified in the following way:
\begin{eqnarray}\label{gravham}
H_{FV}(\hat{{\boldsymbol{p}}})  &=&  \left(\sigma_3 + i \sigma_2 \right)\left[ \left(1+4\phi\right)\frac{\hat{{\boldsymbol{p}}}^2}{2m}  \ + \  m\phi\right] \nonumber \\[2mm] &&+\  \sigma_3 m \, .
\end{eqnarray} 
Consequently, when we rotate in $H_{FV}$ 
from the diagonal mass matrix (with masses $m_1$ and $m_2$) to the  flavor 
mass matrix $\mathbb{M}$, Eq.~\eqref{B34a} becomes
\begin{eqnarray}
&&\mbox{\hspace{-10mm}}i \partial_t \Phi_{_{\rm{I}}} = \left(\sigma_3 + i \sigma_2 \right)\Bigl[\left(1+4\phi\right) \frac{\hat{{\boldsymbol{p}}}^2}{2D}
\ \! \left(m_{_{\rm{II}}}\Phi_{_{\rm{I}}} - m_{_{\rm{I,II}}}\Phi_{_{\rm{II}}} \right)\nonumber \\[2mm]  &&\mbox{\hspace{-8mm}}+ \left(m_{_{\rm{I}}}\Phi_{_{\rm{I}}} + m_{_{\rm{I,II}}}\Phi_{_{\rm{II}}}\right)\phi\Bigr] \ + \ \sigma_3  \ \! \left(m_{_{\rm{I}}}\Phi_{_{\rm{I}}} +  m_{_{\rm{I,II}}}\Phi_{_{\rm{II}}}   \right), \label{gravstep1}
\end{eqnarray}
and similarly for $\Phi_{_{\rm{II}}}$.

Following the analysis of the previous Section, we pursue our argument in momentum representation, in which the large and small components of the particle $\mathrm{I}$ satisfy the following equations:
\begin{eqnarray}
&&\mbox{\hspace{-3mm}}E_{_{\rm{I}}} \Phi_{_{\rm{I}}}^L\nonumber \\[0mm] &&=  \left(1+4\phi\right) \frac{{{\boldsymbol{p}}}^2}{2D}\left(m_{_{\rm{II}}}\Phi_{_{\rm{I}}}^L  + m_{_{\rm{II}}}\Phi_{_{\rm{I}}}^S - m_{_{\rm{I,II}}}\Phi_{_{\rm{II}}}^L - m_{_{\rm{I,II}}}\Phi_{_{\rm{II}}}^S \right)\nonumber \\[2mm]
&&~~~+  \left(m_{_{\rm{I}}}\Phi_{_{\rm{I}}}^L  + m_{_{\rm{I}}}\Phi_{_{\rm{I}}}^S + m_{_{\rm{I,II}}}\Phi_{_{\rm{II}}}^L + m_{_{\rm{I,II}}}\Phi_{_{\rm{II}}}^S \right)\phi\nonumber\\[2mm]
&&~~~+  m_{_{\rm{I}}} \Phi_{_{\rm{I}}}^L + m_{_{\rm{I,II}}} \Phi_{_{\rm{II}}}^L\, , \label{gravstep2}
\end{eqnarray}
\begin{eqnarray}
&&\mbox{\hspace{-3mm}}E_{_{\rm{I}}} \Phi_{_{\rm{I}}}^S\nonumber \\[0mm] &&= \left(1+4\phi\right) \frac{{{\boldsymbol{p}}}^2}{2D}\left(m_{_{\rm{I,II}}}\Phi_{_{\rm{II}}}^L  + m_{_{\rm{I,II}}}\Phi_{_{\rm{II}}}^S - m_{_{\rm{II}}}\Phi_{_{\rm{I}}}^L - m_{_{\rm{II}}}\Phi_{_{\rm{I}}}^S \right)\nonumber \\[2mm]
&&~~~-  \left(m_{_{\rm{I}}}\Phi_{_{\rm{I}}}^L  + m_{_{\rm{I}}}\Phi_{_{\rm{I}}}^S + m_{_{\rm{I,II}}}\Phi_{_{\rm{II}}}^L + m_{_{\rm{I,II}}}\Phi_{_{\rm{II}}}^S \right)\phi\nonumber
\\[2mm]
&&~~~-  m_{_{\rm{I}}} \Phi_{_{\rm{I}}}^S - m_{_{\rm{I,II}}} \Phi_{_{\rm{II}}}^S\, .
\label{gravstep3}
\end{eqnarray}
We obtain the non-relativistic limit by setting $E^{N\!R}_{_{\rm{I}}} \equiv E_{_{\rm{I}}} - m_{_{\rm{I}}}$ and assuming that $E_{_{\rm{I}}} + m_{_{\rm{I}}} \approx 2m_{_{\rm{I}}}$. With this, we can write
\begin{eqnarray}
&&\mbox{\hspace{-3mm}}E^{N\!R}_{_{\rm{I}}} \Phi_{_{\rm{I}}}^L\nonumber \\ &&= \ \left(1+4\phi\right) \frac{{{\boldsymbol{p}}}^2}{2D}\left[m_{_{\rm{II}}}\left(\Phi_{_{\rm{I}}}^L  + \Phi_{_{\rm{I}}}^S\right) - m_{_{\rm{I,II}}}\left(\Phi_{_{\rm{II}}}^L + \Phi_{_{\rm{II}}}^S\right) \right]\nonumber \\[2mm]
&&~~~+ \ \left[m_{_{\rm{I}}}\left(\Phi_{_{\rm{I}}}^L  + \Phi_{_{\rm{I}}}^S\right) + m_{_{\rm{I,II}}}\left(\Phi_{_{\rm{II}}}^L + \Phi_{_{\rm{II}}}^S\right) \right]\phi\nonumber\\[2mm]
&&~~~+ \ m_{_{\rm{I,II}}} \Phi_{_{\rm{II}}}^L\, ,
\label{gravstep4}
\end{eqnarray}
and
\begin{eqnarray}
&&\mbox{\hspace{-3mm}} \Phi_{_{\rm{I}}}^S \nonumber \\
&&= \ \left(1+4\phi\right)\frac{{{\boldsymbol{p}}}^2}{4Dm_{_{\rm{I}}} }\left[m_{_{\rm{I,II}}}(\Phi_{_{\rm{II}}}^L  + \Phi_{_{\rm{II}}}^S) - m_{_{\rm{II}}}(\Phi_{_{\rm{I}}}^L + \Phi_{_{\rm{I}}}^S) \right]\nonumber \\[2mm]
&&~~~- \left[m_{_{\rm{I}}}\left(\Phi_{_{\rm{I}}}^L  + \Phi_{_{\rm{I}}}^S\right) + m_{_{\rm{I,II}}}\left(\Phi_{_{\rm{II}}}^L + \Phi_{_{\rm{II}}}^S\right) \right]\frac{\phi}{2m_{_{\rm{I}}}}\nonumber\\[2mm]
&&~~~- \ \ \frac{m_{_{\rm{I,II}}}}{2m_{_{\rm{I}}}} \ \!\Phi_{_{\rm{II}}}^S\, .
\label{gravstep5}
\end{eqnarray}
In analogy with the previous Section, we now resolve $\Phi_{_{\rm{I}}}^S$ and  $\Phi_{_{\rm{II}}}^S$ in terms of 
$\Phi_{_{\rm{I}}}^L$  and $\Phi_{_{\rm{II}}}^L$. This allows  
to cast Eq.~(\ref{gravstep4}) for $\Phi_{_{\rm{I}}}^L$ (and the analogous equation for $\Phi_{_{\rm{II}}}^L$) in terms of large components only. If post-Newtonian corrections of the order $\mathcal{O}(\boldsymbol{p}\phi)$ are neglected (as their explicit form is irrelevant for the identification of $m_i$ and $m_g$), we obtain after a simple algebra the non-relativistic, Schr\"odinger-like equation for $\Phi_{_{\rm{I}}}^L$ in the regime $m_{_{\rm{I}}}, m_{_{\rm{II}}}\gg|{\boldsymbol{p}}|$, which turns out to be
\begin{eqnarray}\nonumber
E^{N\!R}_{_{\rm{I}}} \Phi_{_{\rm{I}}}^L&=&\left(\frac{{\boldsymbol{p}}^2}{2m_{i,{_{\rm{I}}}} } + m_{_{\rm{I}}}\phi\right) \Phi_{_{\rm{I}}}^L\\[2mm]
&+&   \left\{m_{_{\rm{I,II}}}\left[1 + \phi -   \frac{{\boldsymbol{p}}^2}{2m_{_{\rm{I}}}m_{i,{_{\rm{II}}}} } \right]\right\}\Phi_{_{\rm{II}}}^L\, . 
\end{eqnarray}
From this, we can immediately deduce that, whilst the inertial mass is still represented by the effective quantity $m_{i,{_{\rm{I}}}}$, the gravitational mass $m_{g,{_{\rm{I}}}} $ must be identified with  $m_{_{\rm{I}}}$, and similarly for $I\leftrightarrow II$. This implies a violation of the WEP, which is completely analogous to the WEP already encountered in the case of neutrino mixing.

\section{Inadequacy of Bargmann's Superselection Rule}

In order to keep our presentation consistent, we will now demonstrate that Bargmann's superselection rule (SSR) is not valid in the present context.
Bargmann's SSR arises as a consequence of demanding Galilean covariance for the Schr\"{o}dinger equation. This, in turn, implies that superposition of states with different masses is forbidden~\cite{Barg} and thus one cannot consistently describe unstable or oscillating particles
at non-relativistic energies~\cite{Jackiw,Khare,Balantekin,Oxman}.



While the impossibility of oscillating particles can be easily deduced in the context of
Galileo transformations, it is in conflict with the principles of relativistic quantum theory.
Indeed, the description of systems where superpositions of states with different mass occur can be carried out without difficulty within relativistic quantum mechanics and quantum field theory, and the significant number of mixed particles observed in high-energy particle physics is evidence of the consistency of such treatment. It is thereby unclear why they should cease to oscillate in the non-relativistic limit.

We propose that this issue can be addressed by recognizing that non-relativistic quantum mechanics can be influenced by relativistic effects, which are not visible or even forbidden when Galileo covariance is strictly enforced.
The appearance of such effects is often proclaimed as non-physical and banished from the general framework via superselection rules.
A particular example of the latter is represented by measurable phase shifts in particle mixing.
%
%
To illustrate our point, we will follow the exposition of Ref.~\cite{Green}. 
For the sake of simplicity, we will employ mixed neutrinos, though our argument can be adapted also to scalar particles with minor adjustments. We start by considering plane-wave
solutions of Eq.~\eqref{en1}, which in position representation acquires the form
\bea\label{e1}\non
\lf(i\partial_0-m_e\ri)\varphi_e \ + \ i{\boldsymbol{\si}}\cdot{\boldsymbol{\nabla}}\chi_e &=& m_{e\mu}\varphi_\mu\,,\\[2mm]
-i{\boldsymbol{\si}}\cdot{\boldsymbol{\nabla}}\varphi_e \ - \ \lf(i\partial_0+m_e\ri)\chi_e &=& m_{e\mu}\chi_\mu\,.
\eea
%
%
%
%
By realizing that, for an observer moving with the particle, the plane-wave phase is
\begin{eqnarray}
k^{\mu}x_{\mu} \ = \ \omega{t} - {\bf k}\cdot {\bf x} \ = \ mc^{2}\tau/\hbar\, ,
\end{eqnarray}
where $\tau$ is the observer's proper time (for future convenience, we have reinstated $c$ and $\hbar$), we can
write for the positive-energy plane waves
\begin{eqnarray}
\psi_e  &=&  \cos\theta  e^{-im_1c^2\tau_1/\hbar} u_1(k)  +  \sin\theta e^{-im_2c^2\tau_1/\hbar} u_2(k)
\nonumber \\[1mm]
&=&  e^{-im_1c^2\tau_1/\hbar} [\cos\theta \ \!u_1(k) + \sin\theta \ \! \tilde{u}_2(k)]\, ,
\label{BI}
\end{eqnarray}
with $\tilde{u}_2 = e^{i(m_1-m_2)c^2\tau_1/\hbar}u_2$. Here, $m_1$ and $m_2$ are masses in the mass basis and $\theta$ is the mixing angle. The flavor masses and mixing term $m_e$, $m_\mu$, $m_{e\mu}$ are related to
$m_1$ and $m_2$ through~\cite{BJV}
\bea\nonumber
m_e&=&m_1\,\mathrm{cos}^2\theta+m_2\,\mathrm{sin}^2\theta\,,\\[2mm]\nonumber
m_\mu&=&m_1\,\mathrm{sin}^2\theta+m_2\,\mathrm{cos}^2\theta\,,\\[2mm]
m_{e\mu}&=&\lf(m_2-m_1\ri)\mathrm{sin}\theta\ \!\mathrm{cos}\theta\,.
\label{mass}
\eea 
Let us now apply on $\psi_e$ a sequence of  transformations {\em \`{a} la} Bargmann~\cite{Barg}, but instead of Galilean boosts we use Lorentz boosts. We start from the original system
$S$ and then perform 4 transformations~\cite{Green}
\begin{eqnarray}
&&{\mbox{Translation by $a$ from  $S$ to $S_I$:}}\nonumber \\
&&\mbox{\hspace{1cm}}~ x \rightarrow x + a = x_I\, ,\nonumber\\
&&\mbox{\hspace{1cm}}~ x_0 = x_{0,I}\, . \nonumber \\[2mm]
&&{\mbox{Boost by $v$ from  $S_I$ to $S_{II}$:}} \nonumber \\
&&\mbox{\hspace{1cm}}~ x_{II} = \gamma(x_I - \beta x_{0,I})\, , \nonumber\\
&&\mbox{\hspace{1cm}}~ x_{0,II} = \gamma(x_{0,I} - \beta  x_I)\, .\nonumber \\[2mm]
&&{\mbox{Translation by $-a$ from  $S_{II}$ to $S_{III}$:}}\nonumber \\
&&\mbox{\hspace{1cm}}~ x_{II} \rightarrow x_{II} - a/\gamma = x_{III}\, , \nonumber\\
&&\mbox{\hspace{1cm}}~ x_{0,II} = x_{0,III}\, . \nonumber \\[2mm]
&& {\mbox{Boost by $-v$ from  $S_{III}$ to $S_{IV}$:}} \nonumber \\
&&\mbox{\hspace{1cm}}~ x = x_{IV} = \gamma(x_{III} + \beta x_{0,III})\, , \nonumber\\
&&\mbox{\hspace{1cm}}~ x_{0,IV} = \gamma(x_{0,III} + \beta  x_{III})\nonumber \\
&&\mbox{\hspace{1.8cm}}~  = x_{0} - \beta a\, .
\end{eqnarray}
Here, $\beta = v/c$ and $\gamma = (1- \beta^2)^{-1/2}$.
After the sequence of transformations $S \rightarrow S_I\rightarrow S_{II} \rightarrow S_{III} \rightarrow S_{IV}$  we end up in the original point ${\bf x}$ but in the Lorentz shifted time $t_{IV} \neq t$. In the version with Galileo boosts, we would have $t=t_{IV}$, and so we would end up in the frame $S_{IV} =S$. From the point of view of the observer who has undertaken the sequence of
above transformations, the mixing (\ref{BI}) reads
\begin{eqnarray}
&&\mbox{\hspace{-11mm}}\psi_e'  \ = \  \cos\theta  e^{-im_1c^2\tau_2/\hbar} u_1  +  \sin\theta e^{-im_2c^2\tau_2/\hbar} u_2
\nonumber \\[1mm]
&&\mbox{\hspace{-6mm}}= \ e^{-im_1c^2\tau_2/\hbar} [\cos\theta \ \!u_1 + \sin\theta \ \! e^{-i\Delta mc^2 \Delta \tau/\hbar} \tilde{u}_2]\, ,
\end{eqnarray}
where  $\Delta m = m_1 - m_2$ and $\Delta \tau =  \tau_1 - \tau_2$ is the difference between proper times of both observers. Note that the momentum bispinors $u_1$ and $u_2$ are not affected by the combined transformation, as the net effect of the sequence of transformations is reflected only in the phase parts.

Due to the lack of simultaneity between the two observers (twin paradox), the two states $\psi_e$ and $\psi_e'$
must be different (i.e., they are not members of the same projective ray in the Hilbert space). The appearance of the  extra relative phase factor $e^{-i\Delta mc^2 \Delta \tau/\hbar}$
is thus not surprising as $\Delta \tau \neq 0$.

In the Galilean framework, the analogous situation looks differently. The sequence of the four transformations gives an identity operation and the presence of a non-relativistic analogue of
the above relative phase is inconsistent with the fact that the state $\psi_e$ should coincide with the state $\psi_e'$ (they lie on the same ray). In fact, such a relative phase does not
have any meaning and is forced to be 1 by proclaiming that the only logically consistent situation is $m_1 = m_2$  (i.e., Bargmann's SSR), implying that no neutrino mixing can take place non-relativistically.

Let us  note that the phase factor $e^{-i\Delta mc^2 \Delta \tau/\hbar}$ will not disappear in the non-relativistic limit, but it will leave an imprint that is independent of $c$.
Indeed
\begin{eqnarray}
\Delta \tau &=& \tau_1 - \tau_2 \ = \ t  - \int_{t_0}^{t_{0,IV}} \sqrt{1 - \frac{v^{2}(t)}{c^2}} \ \! dt \nonumber \\[2mm]
&=& t - \int_{t_{0,I}}^{t_{0,II}} \sqrt{1 - \frac{v^{2}}{c^2}} \ \! dt \ + \ \int_{t_{0,III}}^{t_{0,IV}} \sqrt{1 - \frac{v^{2}}{c^2}}\ \! dt  \nonumber \\[2mm]
&=& t - \sqrt{1 - \frac{v^{2}}{c^2}} \ \!t \ \ \stackrel{NR}{\rightarrow} \ \ \frac{v^2}{2 c^2} t_{NR}  \ = \  \frac{v a}{ c^2}\, ,
\end{eqnarray}
where $ct = x_{0,IV} - x_{0}$ and $t_{NR} = [(a + a/\gamma)/v]_{NR} = 2a/v$. Hence, we obtain that
\begin{eqnarray}
e^{-i\Delta mc^2 \Delta \tau/\hbar} \  \ \stackrel{NR}{\rightarrow} \  \ e^{-i\Delta m v a/\hbar}\, .
\end{eqnarray}
%
%
This phase factor, though not depending on $c$, has no basis in the Galileo transformations (where the concept of proper time is meaningless) and is erroneously seen as non-physical and removed via SSR.

In short, the non-relativistic limit of superpositions of states with different
mass can comfortably accommodate a relative phase that is otherwise problematic
from the point of view of Galileo transformations. So, neutrino mixing and ensuing oscillations do not pose
any conceptual difficulties in the non-relativistic limit, and certainly they are
not prohibited by Bargmann's SSR.
Similar considerations hold true also for scalar particle in FVR. 
The only difference is that for spinless particles $\psi_e \rightarrow \Phi_{_{\rm{I}}}$, $m_e, m_\mu, m_{e\mu} \rightarrow  m_{_{\rm{I}}}, m_{_{\rm{II}}}, m_{_{\rm{I,II}}}$ and $\theta \rightarrow \tilde{\theta}$, where $\tilde{\theta}$ is the mixing angle through which we must rotate the diagonal mass-matrix to obtain $\mathbb{M}$ in (\ref{20.cv}).

\section{Conclusions}

In this Letter, we have investigated the non-relativistic limit of the Klein--Gordon equation for mixed scalar particles.   
To mimic our treatment for spin-1/2 particles outlined in
Ref.~\cite{BJLP}  (and summarized in Sec.~II), we have employed the Feshbach--Villars representation, according to which the wave function of a spinless particle becomes a two-component object and the equation of motion 
is of the first order in time. Within this setting, we have demonstrated that the resulting Schr\"odinger-like equation predicts an effective inertial masses, which does not coincide with eigenvalues of the mass matrix in the relativistic regime. In particular, the ensuing low-energy inertial masses non-trivially depend on a mixing term.

We have also shown that, when a weak external gravitational field is taken into account, the resulting gravitational masses remain unchanged in the non-relativistic limit, thereby giving rise to a violation of WEP. Interestingly, the rate of this violation is
identical to the one encountered in the framework of spinor flavor mixing, the only difference being an overall numerical factor which is associated to the spin of the considered particle. 
Finally, we have stressed that the non-relativistic limit for superpositions of states with different mass does not produce any inconsistency in non-relativistic quantum mechanics, because in this scenario Bargmann's SSR stemming from the enforcement of Galilean covariance 
is not applicable. 

Finally, it is important to note that, unlike the case in neutrino physics,  the current model of meson oscillations is inevitably an effective model since mesons are not fundamental particles.  Indeed, a full-fledged treatment should involve quarks, whose mixing properties are encoded in the Cabibbo--Kobayashi--Maskawa matrix~\cite{ckm}. However, such an analysis would pose a series of problems, the majority of which are related to the fact that mesons are made up of quarks tied together by the strong force. In order to properly understand the issue at high enough energies, it is necessary to use quantum field theory instead of first quantization. For energies lower than  $m_1$ and $m_2$,  it is reasonable to assume that our first-quantized analysis should be viable.

\section*{Acknowledgments}
L.P. acknowledges support by the COST Action CA18108 and MUR (Ministero dell'Universit\`a e della Ricerca) via the project PRIN 2017 ``Taming complexity via QUantum Strategies: a Hybrid Integrated Photonic approach'' (QUSHIP) Id. 2017SRNBRK and is grateful to the ``Angelo Della Riccia'' foundation for the awarded fellowship received to support the study at Universit\"at Ulm.
P.J. was in part supported by the FNSPE CTU grant RVO14000.



\begin{thebibliography}{99}

\bibitem{ep}
C.M.~Will, Living Rev. Rel. \textbf{9}, 3 (2006); Living Rev. Rel. \textbf{17}, 4 (2014).

\bibitem{ep2}
C.M.~Will, \emph{Theory and Experiment in Gravitational Physics} (Cambridge, Cambridge
University Press, 2018).

\bibitem{ep3}
 E.~Di Casola, S.~Liberati and S.~Sonego, Am. J. Phys. \textbf{83}, 39 (2015).


\bibitem{ep4}
M.~Blasone, S.~Capozziello, G.~Lambiase and L.~Petruzziello,
Eur. Phys. J. Plus \textbf{134}, 169 (2019).

\bibitem{ep5}
M.~Blasone, S.~Capozziello, G.~Lambiase and L.~Petruzziello,
Int. J. Geom. Meth. Mod. Phys. \textbf{19}, 2250055 (2022).

\bibitem{wep6}
G.M.~Tino, L.~Cacciapuoti, S.~Capozziello, G.~Lambiase and F.~Sorrentino,
Prog. Part. Nucl. Phys. \textbf{112}, 103772 (2020).

\bibitem{sck}
  I.I.~Shapiro, C.C.~Counselman and R.W.~King,
  Phys.\ Rev.\ Lett.\  {\bf 36}, 555 (1976).	
	
\bibitem{nmf}	
 T.M.~Niebauer, M.P.~Mchugh and J.E.~Faller,
  Phys.\ Rev.\ Lett.\  {\bf 59}, 609 (1987).
	
\bibitem{sah}	
C.W.~Stubbs, E.G.~Adelberger, B.R.~Heckel, W.F.~Rogers, H.E.~Swanson, R.~Watanabe, J.H.~Gundlach and F.J.~Raab,
  Phys.\ Rev.\ Lett.\  {\bf 62}, 609 (1989).
	
\bibitem{ash}
  E.G.~Adelberger, C.W.~Stubbs, B.R.~Heckel, Y.~Su, H.E.~Swanson, G.~Smith, J.H.~Gundlach and W.F.~Rogers,
  Phys.\ Rev.\ D {\bf 42}, 3267 (1990).
	
\bibitem{bha}	
  S.~Baessler, B.R.~Heckel, E.G.~Adelberger, J.H.~Gundlach, U.~Schmidt and H.E.~Swanson,
  Phys.\ Rev.\ Lett.\  {\bf 83}, 3585 (1999).
	
\bibitem{nprd}
L.~Buoninfante, G.G.~Luciano, L.~Petruzziello and L.~Smaldone,
  Phys.\ Rev.\ D {\bf 101}, 024016 (2020).	
	
\bibitem{ijmpd}
G.G.~Luciano and L.~Petruzziello,
Int. J. Mod. Phys. D \textbf{29}, 2043002 (2020).
	
\bibitem{gn}
 A.M.~Gago, H.~Nunokawa and R.~Zukanovich Funchal,
  Nucl.\ Phys.\ Proc.\ Suppl.\  {\bf 100}, 68 (2001).	
	
\bibitem{hl}
 J.T.~Pantaleone, A.~Halprin and C.N.~Leung,
  Phys.\ Rev.\ D {\bf 47}, R4199 (1993);
  A.~Halprin, C.N.~Leung and J.T.~Pantaleone,
  Phys.\ Rev.\ D {\bf 53}, 5365 (1996).
  
\bibitem{BJLP}
M.~Blasone, P.~Jizba, G~Lambiase and L.~Petruzziello, Phys. Let. B \textbf{811}, 135883 (2020). 

\bibitem{dmes}
R.~Aaij \emph{et al.} [LHCb], Phys. Rev. Lett. \textbf{127}, 111801 (2021).

\bibitem{mam}
R.~Aaij \emph{et al.} [LHCb],
Phys. Lett. B \textbf{728}, 607 (2014).

\bibitem{mam2}
R.~Aaij \emph{et al.} [LHCb],
Nature Phys. \textbf{18}, 1 (2022).

\bibitem{FV}
H.~Feshbach and F.~Villars, Rev. Mod. Phys. \textbf{30}, 24 (1958).

\bibitem{JK}
P.~Jizba and H.~Kleinert, Phys. Rev. D \textbf{82}, 085016 (2010).

\bibitem{BD}
N.D.~Birrel and P.C.W.~Davies, {\em Quantum Fields in Curved
Space}, (Cambridge University Press, Cambridge, England, 1994).

\bibitem{fvgrav}
A.~Accioly and H.~Blas,
Phys. Rev. D \textbf{66}, 067501 (2002).

\bibitem{GP}
A.~Grib and E.~Poberii, Helv. Phys. Acta {\bf 68}, 380 (1995).

\bibitem{Barg}
  V.~Bargmann, Ann. Math. \textbf{59}, 1 (1954).

\bibitem{BJV} M.~Blasone, P.~Jizba and G.~Vitiello, {\em Quantum Field Theory and its Macroscopic Manifestations}, (World Scientific \& ICP, London, 2010).

\bibitem{Feyn}	
R.P.~Feynman, R.B.~Leighton and M.L.~Sands, \emph{The Feynman lectures on physics, Vol.3}, Reading, Mass: Addison-Wesley Pub. Co. (1963).

\bibitem{Jackiw}
R. Jackiw and S. Templeton, Phys. Rev. D \textbf{23}, 2291 (1981). 

\bibitem{Khare}
A. Khare and J. Maharana, Phys. Lett. B \textbf{209}, 468 (1988). 

\bibitem{Balantekin}
A.B.~Balantekin and G.M.~Fuller, Phys. Rev. D \textbf{103}, 113003 (2021)

\bibitem{Oxman}
L.E.~Oxman and A.L.~de Queiroz, Ann. Phys. \textbf{426}, 168346 (2021).  

\bibitem{Green}
  D.M.~Greenberger, Phys. Rev. Lett. \textbf{87}, 100405 (2001).
  
\bibitem{ckm}
N.~Cabibbo,
Phys. Rev. Lett. \textbf{10}, 531 (1963);
M.~Kobayashi and T.~Maskawa,
Prog. Theor. Phys. \textbf{49}, 652 (1973).

\end{thebibliography}
\end{document}